\documentclass[12pt]{article}
\usepackage[dvips]{graphicx}
\usepackage{pdproc}

  \makeatletter 
  \def\@cite#1{[#1]} 
  \makeatother    
\begin{document}

\renewcommand{\thefootnote}{\alph{footnote}}

\newcommand{\hx}{\hat{x}}
\newcommand{\hxm}{\hat{x}^{\mu}}
\newcommand{\hxn}{\hat{x}^{\nu}}
\newcommand{\tmn}{\theta^{\mu\nu}}
\newcommand{\hD}{\hat{\Delta}}
\newcommand{\mo}{\mathcal{O}}
\newcommand{\mL}{\mathcal{L}}
\newcommand{\pd}{\partial}
\newcommand{\vp}{\varphi}
\newcommand{\bR}{\mathbb{R}}
\newcommand{\nn}{\nonumber}
\newcommand{\e}{{\rm e}}
\newcommand{\tr}{\rm Tr}
\newcommand{\del}{\delta}
\newcommand{\ra}{\rangle}
\newcommand{\la}{\langle}
\newcommand{\rar}{\rightarrow}
\newcommand{\lar}{\leftarrow}
\newcommand{\lra}{\longrightarrow}
\newcommand{\lla}{\longleftarrow}
\newcommand{\pdm}{\partial_{\mu}}
\newcommand{\pdn}{\partial_{\nu}}
\newcommand{\fmn}{F_{\mu\nu}}
\newcommand{\am}{A_{\mu}}
\newcommand{\an}{A_{\nu}}
\newcommand{\dg}{\dagger}
\newcommand{\db}{\delta_{\rm B}}
\newcommand{\bdb}{\bar{\delta}_{\rm B}}
\newcommand{\bx}{\bar{x}}
\newcommand{\fx}{x_{f}}
\newcommand{\ix}{x_{i}}
\newcommand{\vph}{\vec{\phi}}
\newcommand{\ph}{\phi}
\newcommand{\hphi}{\hat{\phi}}
\newcommand{\al}{\alpha}
\newcommand{\be}{\beta}
\newcommand{\fr}{\frac}
\newcommand{\bfx}{{\bf x}}
\newcommand{\lam}{\lambda}
\newcommand{\lag}{{\mathcal L}}
\newcommand{\cpn}{\mathbb{C}P^N}
\newcommand{\bfn}{{\bf n}}
\newcommand{\Dm}{D_{\mu}}
\newcommand{\Dn}{D_{\nu}}
\newcommand{\ep}{\epsilon}
\newcommand{\bfw}{{\bf w}}
\newcommand{\epmn}{\epsilon_{\mu\nu}}
\newcommand{\cp}{\mathbb{C}P}
\newcommand{\bz}{\bar{z}}
\newcommand{\bpd}{\bar{\pd}}
\newcommand{\sig}{\sigma}
\newcommand{\emn}{\eta_{\mu\nu}}
\newcommand{\iemn}{\eta^{\mu\nu}}
\newcommand{\mN}{\mathcal{N}}
\newcommand{\slD}{\not{\! \mbox{D}}}
\newcommand{\da}{a^{\dagger}}
\renewcommand{\th}{\theta}
\newcommand{\xmu}{x^{\mu}}
\newcommand{\xnu}{x^{\nu}}
\newcommand{\bth}{\bar{\theta}}
\newcommand{\delb}{\delta_{\rm B}}
\newcommand{\bdelb}{\bar{\delta}_{\rm B}}
\newcommand{\red}{\textcolor{red}}
\newcommand{\green}{\textcolor{green}}
\newcommand{\magenta}{\textcolor{magenta}}
\newcommand{\cyan}{\textcolor{cyan}}
\newcommand{\blue}{\textcolor{blue}}
\newcommand{\black}{\textcolor{black}}
\newcommand{\tg}{\tilde{\gamma}}
\newcommand{\s}{\scriptscriptstyle}
\newcommand{\hu}{\hat{u}}
\newcommand{\hv}{\hat{v}}
\newcommand{\I}{\bf 1}
\newcommand{\up}{\uparrow}
\newcommand{\down}{\downarrow}
\newcommand{\Rb}{\mathbb{R}}
\newcommand{\Cb}{\mathbb{C}}
\newcommand{\non}{\nonumber\\}
\newcommand{\dis}{\displaystyle}
\newcommand{\1}{\mathbb I}
\newcommand{\CC}{\mathcal C}
\newcommand{\CM}{\mathcal M}
\newcommand{\CO}{\mathcal O}
\newcommand{\CP}{\mathcal P}
\renewcommand{\baselinestretch}{1.4}
\newcommand\Tr{{\rm Tr}}

\title{
D-branes of Covariant AdS Superstrings
\footnote
{Talk was presented by K.\ Y.\ 
at {\it SUSY 2004: 
The 12th International Conference on Supersymmetry
and Unification of Fundamental Interactions}, 
held at Epochal Tsukuba, Tsukuba, Japan, 
June 17-23, 2004. To appear in the Proceedings.
} 
\\-- An Overview -- }

\author{MAKOTO SAKAGUCHI}

\address{
Osaka City University Advanced Mathematical Institute (OCAMI), \\
3-3-138, Sugimoto, Sumiyoshi-ku, Osaka 558-8585, Japan. 
\\ {\rm E-mail: msakaguc@sci.osaka-cu.ac.jp}}

\author{KENTAROH YOSHIDA}

\address{ 
Theory Division, High Energy Accelerator Research 
Organization (KEK), \\
1-1 Oho, Tsukuba, Ibaraki 305-0801, Japan.
\\ {\rm E-mail: kyoshida@post.kek.jp}}

\abstract{ We briefly review a covariant analysis of D-branes of type
IIB superstring on the AdS$_5$$\times$S$^5$ background from 
the $\kappa$-invariance of the Green-Schwarz string action. The 
possible configurations of D-branes preserving half of supersymmetries
are classified in both cases of AdS$_5$$\times$S$^5$ and the pp-wave
background.  }

\normalsize\baselineskip=15pt

\section{Introduction} 

D-brane is an important key ingredient in studies of non-perturbative
aspects of superstring theories, and it is a recent interest to study
D-branes on curved backgrounds.  In particular, those on pp-wave
backgrounds \cite{BFHP1} have been well studied,
since the Green-Schwarz 
strings on pp-waves are exactly solvable 
in light-cone gauge \cite{M} and so one can study 
them 
directly by quantizing the theories \cite{BP,DP,SMT,BGG}. 

Covariant studies of D-branes in type IIB and IIA strings on
pp-waves were discussed in \cite{BPZ} and \cite{HPS}, respectively, by
following the method of Lambert and West \cite{LW}.  Motivated by these
developments, we have carried out a covariant analysis of D-branes of 
type IIB string on the AdS$_5$$\times$S$^5$ background
\cite{SaYo4,sfull}, 
by using the Green-Schwarz action obtained by Metsaev and Tseytlin 
\cite{MT:action}.   
The possible 1/2 supersymmetric (SUSY) D-brane configurations have been
classified. This result is consistent to that of brane probe
analysis in \cite{SMT}.  In addition, Penrose limits \cite{P,BFHP2,HKS} 
of D-branes on the AdS$_5$$\times$S$^5$ give 
possible D-brane configurations in the type IIB pp-wave background.  

On the other hand, by employing 
the methods of \cite{EMM}, the covariant analysis 
is also applicable to open supermembranes on the pp-wave
\cite{SY1,SaYo:pp} and AdS$_{4/7}$$\times$S$^{7/4}$ \cite{SaYo:ads,mfull} 
backgrounds. These results are related via Penrose limit and are also
consistent to the brane probe analysis in eleven dimensions
\cite{Kim-Yee}. 

We will briefly review the classification of D-branes on the
AdS$_5$$\times$S$^5$ preserving 
half of supersymmetries, and 
discuss the Penrose limit of them.  

\newpage 

\section{The action of type IIB string on the AdS$_5$$\times$S$^5$}
First of all, the action of AdS string we consider is written as 
\cite{MT:action} 
\begin{eqnarray}  
\label{S} 
S = \int\!\!d^2\sigma\,\Bigl[\mathcal{L}_{\rm NG} + 
\mathcal{L}_{\rm WZ}\Bigr]\,, \quad \mathcal{L}_{\rm NG} = -\sqrt{-g(X,\th)}
\,. 
\end{eqnarray}
The Nambu-Goto part of this Lagrangian is represented in terms of the 
induced metric $g_{ij}$, which is given by (For notation and convention,
see \cite{SaYo4}) 
\begin{eqnarray}
g_{ij} = E_i^ME^N_j G_{MN} = E_i^AE_j^B\eta_{AB}\,, \quad 
g = \det g_{ij}, \quad 
E^A_i=\partial_iZ^{\hat M}E_{\hat M}^A\,,
\end{eqnarray}
where $Z^{\hat M}=(X^M,\theta^{\bar \alpha})$ and $E_{\hat M}^A$ are
supervielbeins of the AdS$_5$$\times$S$^5$ background.  For D-strings,
$g$ is replaced with $\det(g_{ij}+{\mathcal F}_{ij})$ where ${\mathcal
F}$ is defined by ${\mathcal F}=dA-B$ with the Born-Infeld $U(1)$ gauge
field $A$ and the pull-back of the NS-NS two-form $B$. The Wess-Zumino
term, which is needed for the $\kappa$-invariance and makes the theory
consistent, is
\begin{eqnarray}
\label{WZ}
\mathcal{L}_{\rm WZ} = -2i\int_0^1
\!\!dt\,\widehat{E}^{A}\bar{\th}\Gamma_A\sigma 
\widehat{E}\,,
\end{eqnarray}
where $\widehat{E}^A \equiv E^A(t\th)$ and $\widehat{E}^{\al}\equiv 
E^{\al}(t\th)$. 
When we consider a fundamental string (F-string), the matrix $\sigma$ 
is given by $\sigma_3$\,. If we consider a D-string, then $\sigma$ is
represented by $\sigma_1$\,. Since we would like to discuss boundary
surfaces for both of 
F-string and D-string, we do not
explicitly fix $\sigma$ in our consideration. 

\section{D-branes from $\kappa$-invariance}

Let us consider D-branes on the AdS$_5$$\times$S$^5$ 
by following the idea of Lambert and West \cite{LW}.
They considered the
D$p$-branes from the $\kappa$-invariance of the Green-Schwarz type IIB 
string in flat space and obtained the standard fact that the value $p$
is odd. Such a constraint comes from the requirement that we
should impose appropriate boundary conditions in order to delete   
the surface terms coming from the $\kappa$-variation and to ensure the
consistency of the theory. 

The idea of Lambert and West can be applicable to non-trivial
backgrounds, including the AdS$_5$$\times$S$^5$
and the pp-wave. In these cases, the boundary conditions 
restrict not only the value $p$ but also the configuration of a D$p$-brane,
and lead to the classification of possible D-branes
\cite{BPZ,HPS,SaYo4,sfull,SaYo:pp,SaYo:ads,mfull}.

\subsection{The classification of 1/2 SUSY D-branes 
on the AdS$_5$$\times$S$^{\,5}$} 

The classification of 1/2 SUSY D-branes on the AdS$_5$$\times$S$^5$ 
\cite{SaYo4} 
was given by considering the vanishing conditions of
the $\kappa$-variation surface terms up to and including the fourth
order in $\theta$. This result is still valid even at full order of
$\th$ \cite{sfull}.   
The result is as follows: 
For the $d=2$ (mod 4) case, where $d$ is the number of Dirichlet
directions, the possible
configurations of D-branes need to satisfy the condition:
\begin{itemize}
 \item The number of Dirichlet directions in the AdS$_5$ coordinates
       $(X^0,\cdots,X^4)$ is even, and the same condition is also satisfied for
       the S$^5$ coordinates $(X^5,\cdots,X^9)$.
\end{itemize}
For the $d=4~({\rm mod}~4)$ case, 
D-branes satisfying the following condition are allowed:  
\begin{itemize}
 \item The number of Dirichlet directions in the AdS$_5$ coordinates
       $(X^0,\cdots,X^4)$ is odd, and the same condition is also satisfied for
       the S$^5$ coordinates $(X^5,\cdots,X^9)$.
\end{itemize}
For a D-brane on the AdS$_5$$\times$S$^5$, the directions to which the
brane world-volume can extend are restricted. All the possible D-brane
configurations at the origin are summarized in Tab.\,\ref{tab1}. When we
consider the D-branes sitting outside the origin, only a D-instanton is
allowed 
as a 1/2 SUSY object. 
\begin{table}[htbp]
\caption{The possible 1/2 SUSY D-branes of F (D)-string on 
the AdS$_5$$\times$S$^5$\,, sitting at the origin.}
 \begin{center}
{\small 
  \begin{tabular}{|c|c|c|c|c|c|}
\hline
D-instanton & D (F)-string  & D3-brane  & D5 (NS5)-brane & D7 
& D9 (NS9)-brane \\
\hline\hline
(0,0) & (0,2),~(2,0) & (1,3),~(3,1) & (2,4),~(4,2)
& (3,5),~(5,3) & absent \\
\hline
  \end{tabular}
}
 \end{center}
\label{tab1}
\end{table}

\subsection{Penrose Limit of D-branes on the AdS$_5$$\times$S$^{\,5}$}

The Penrose limit \cite{P} of the AdS$_5$$\times$S$^5$ background leads
to the maximally supersymmetric pp-wave background \cite{BFHP2}. 
We may consider the Penrose limit of our classification result presented
in the previous subsection. Then we can classify the possible D-branes
on the pp-wave, including the well-known results in the light-cone
analysis of the pp-wave string \cite{BP,DP,BGG} (For the detail, 
see our work \cite{SaYo4}). 
The result is summarized in 
Tab.\,\ref{penrose:tab},  
which reveals the AdS origin of D-branes on the pp-wave.
It is also consistent with the brane probe analysis
\cite{SMT}. 
Notably, we can see why 1/2 SUSY D-strings do not appear in the 
light-cone analysis.

\begin{table}[htbp]
\caption{Penrose limit of D-branes on the AdS$_5$$\times$S$^5$.}
\begin{center}
{\small
\begin{tabular}{cc}
$
   \begin{array}{cccc}
\hline 
\multicolumn{4}{c}{\mbox{D7-brane}}
\\
\hline
\multicolumn{2}{c}{(3,5)~~}&
\multicolumn{2}{c}{(5,3)~~}   \\  
D^2 \swarrow & \searrow N^2 &  D^2 
\swarrow & \searrow N^2  \\
- & (+,-;2,4)  & - 
& (+,-;4,2) \\
\hline
   \end{array}
$ 
 
\quad &  \quad 

$ 
   \begin{array}{cccc}
\hline 
 \multicolumn{4}{c}{\mbox{D5 (NS5)-brane}} 
\\
\hline
\multicolumn{2}{c}{(2,4) \quad } &\multicolumn{2}{c}{(4,2) \quad }  \\  
D^2 \swarrow  & \searrow N^2 & D^2 
\swarrow & \searrow N^2  \\
(2,4) & (+,-;1,3) & (4,2) &
 (+,-;3,1) \\
\hline
   \end{array}
$ 
\\ 
\quad & \quad  \\
$ 
   \begin{array}{cccc}
\hline 
 \multicolumn{4}{c}{\mbox{D3-brane}}  
\\
\hline
\multicolumn{2}{c}{(1,3)~~}&\multicolumn{2}{c}{(3,1)~~}  \\  
D^2 \swarrow  & \searrow N^2 & D^2 
\swarrow & \searrow N^2  \\
(1,3) & (+,-;0,2) & (3,1) & (+,-;2,0)  \\
\hline
   \end{array}
$ 
\quad &  \quad 
$ 
   \begin{array}{cccc}
\hline 
 \multicolumn{4}{c}{\mbox{D (F)-string}} 
\\
\hline
\multicolumn{2}{c}{(0,2)~ \quad } & \multicolumn{2}{c}{(2,0)~ \quad } \\  
D^2 \swarrow  & \searrow N^2 & D^2 
\swarrow & \searrow N^2  \\
(0,2) & \quad \qquad -  \quad    
& (2,0) & \quad\qquad  - \quad   \\
\hline
   \end{array}
$ 
\end{tabular}
\mbox{$-$:{\scriptsize We cannot take this boundary condition.}} 
}
\end{center}
\label{penrose:tab}
\end{table}

\newpage

\section{Acknowledgments}

The work of M.~S.\ is supported by the 21 COE program
``Constitution of wide-angle mathematical basis focused on knots''. 
The work of K.~Y.\ is supported in part by JSPS Research Fellowships for
Young Scientists.

\bibliographystyle{plain}

\end{document}